\documentstyle[prc,aps,psfig,twocolumn]{revtex}
\begin{document}

\twocolumn[
\hsize\textwidth\columnwidth\hsize\csname @twocolumnfalse\endcsname

\draft

\title{Existence of Dynamical Scaling in the Temporal Signal of Time Projection Chamber}

\author{A. K. Mohanty, D. Favretto, F. Carminati and K. Safarik}
\address{CERN,\\
1211, Geneva, Switzerland}

\maketitle

\begin{abstract}
The temporal signals from a large gas detector may show dynamical scaling due to many
correlated space points created by the charged particles while passing through the tracking medium.
This has been demonstrated through simulation using realistic parameters of a Time Projection Chamber
(TPC) being fabricated to be used in ALICE collider experiment at CERN. An interesting aspect of this
dynamical behavior is the existence of an universal scaling which does not depend on the multiplicity
of the collision. This aspect can be utilised further to study physics at the device level and also 
for the online monitoring
of certain physical observables including electronics noise which are a 
few crucial parameters for the optimal
TPC performance. 

\end{abstract}

%\pacs{PACS numbers:25.70.Jj,24.Eq,25.60.-t,25.70.Gn}

]
\narrowtext
Dynamical scaling refers here to powerlaw distribution or a powerlaw with an exponential cut-off that
describes certain correlation phenomena in the dynamical systems distinctly being different from the
random statistical processes. In the thermodynamical context, it describes a critical phenomena  associated with
the phase transition \cite{julio} while in many other complex systems it corresponds to the so called self
organised criticality \cite{bak}. Although a powerlaw distribution $(f\sim x^{-\beta})$ is a common dynamical feature,
a powerlaw with an additional exponential $(f\sim x^{-\beta} e^{-\alpha x})$ makes the distribution
normalizable for all values of $\beta$ (where $\alpha$ and $\beta$ are positive constants) 
and also many real world systems like World Wide Web and social networks
show this cut-off \cite{albert}. In this paper, we show the presence of both type of scalings, a powerlaw and a powerlaw
with exponential in the temporal signals (comprising of time gap and bunch length distributions)
of a large Time Projection Chamber (TPC) which is a type of gas detector used for
three dimensional tracking of the charged particles passing through it during a particle physics experiment. 
The time gap distribution along the drift direction shows a dynamical scaling which is independent of the multiplicity
of the collisions above a critical value. While this scaling
behavior itself is an interesting aspect to investigate the correlation phenomena in a gas detector 
at the device level, it can be further utilised for online monitoring of certain important TPC
parameters including the constant rise in electronics noise level due to long exposure to radiation.

The TPC is the main tracking device of the ALICE (A Large Ion Collider Experiment) \cite{alice} at the
Large Hadron Collider (LHC) at CERN optimized for the study of heavy ion collisions at a 
centre of mass energy $\sim5.5$ ATeV.
It is a large gas filled detector of cylindrical design with an inner radius of about 80 cm, an outer radius of about
$250$ cm, and an over all length in the beam direction of $500$ cm. A charged particle passing through the gas volume
creates electrons by ionization. The electrons drift in the electric field towards the read out chambers (multiwire
proportional counters with more than $550000$ cathode pads read out located at the two end-caps of the TPC cylinder) where
they are amplified in the field of the sense wires by a factor of several $10^4$. This signal is coupled to the read out pads
which are on ground potential and at a few milimeters distance behind the sense wire-plane. The detail aspect of the
design and simulation results can be found in ref \cite{tpc}. For simulation, the charged particles (mostly pions) with
different multiplicities are generated using HIJING parametrization corresponding to $Pb+Pb$ collisons at 5.5 ATeV. The
simulation is carried out using a microscopic simulator \cite{kow} incorporated in ALIROOT which is a GEANT3.21 and ROOT
based simulation package used by the ALICE collaboration \cite{aliroot}. 

While we refer to \cite{kow} for deatil, in the following we
briefly mention the salient features of the simulation.  The ionization in the gas proceeds in two stages. Firstly,
the electromagnetic interactions of the primary particles with the TPC gas (90$\%$ Ne and 10$\%$ CO$_2$) lead to the
release of primary electrons with a statistics that follows a Poisson distribution. Thus, the distance $S$ between two
successive collisons
leading to primary ionization can be simulated through an exponential distribution given by $\exp(-S/D)/D$ 
where $D=(N_{prim} f)^{-1}$ is the mean distance between primary ionizations, $N_{prim}$ is the 
number of primary electrons per cm produced by a Minimum Ionizing Particle (MIP) and $f$ is the Bethe-Bloch
curve. At sufficient kinetic energy, the primary electrons produce secondaries creating an electron cluster with a total
number of electrons given by $N_{tot}=(E_{tot}-I_{pot})W_i+1$ where $E_{tot}$ is the energy loss in a given collision,
$W_i$ is the effective energy required to produce an electron-ion pair and $I_{pot}$ is the first ionization potential.
The electron cluster which is assumed to be point like undergoes diffusion while drifting towards the end-cap
which is described by a three dimensional Gaussian with widths $\delta_T=D_T\sqrt{L_d}$ and $\delta_L=D_L\sqrt{L_d}$
where $D_T$ and $D_L$ are transverse and longitudinal diffusion constants and $L_d$ is the total drift distance.
An electron arriving at the anode wire creates an avalanche which induces a charge on the pad plane. The time
signal is obtained by folding the avalanche with the shaping function of the pre-amplifier/shaper with a shaping time
$\sim 200$ ns which is a compromise between the need for achieving a high signal to noise ratio and for avoiding
overlap of successive signals.
This signal is sampled with a frequency $\sim 5.66~MHz$ which divides the total drift time of $88~\mu s$ into about
$500$ time bins. The microscopic simulator also takes into account the electron loss in the drift gas due to presence of 
electron negative gas like O$_2$ and also ${\bf ExB}$ effect near the anode wires. The signal is digitized
by a $10$ bit A/D converter that generates Gaussian random noise with r.m.s. about $1000~e$.
Finally, the digitized data is processed and formatted by an Application Specific Integrated Circuit (ASIC) called
ALTRO (ALICE TPC Read Out) \cite{musa}.  A few typical parameters which are used in the simulations are given in table I.  
Although, we concentrate only on the TPC data, all component of the ALICE detectors as well as all passive materials are
included in the simulation so as to create a situation close to the real experiment.

\begin{figure}
\centerline{\hbox{
\psfig{figure=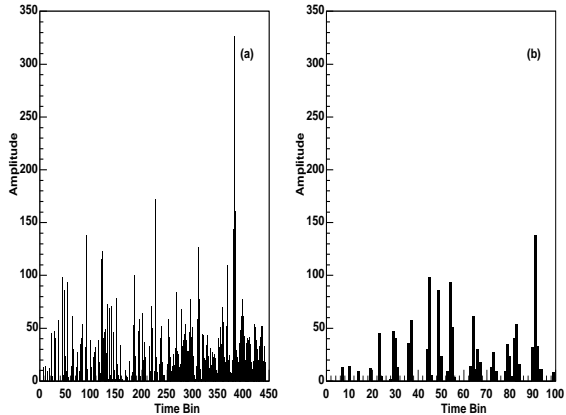,width=3.0in,height=2.6in}}}
\caption{(a) The amplitude versus time bin corresponding to a single pad.
 (b) An expanded
view over 100 time bins.}
\end{figure}

In ALTRO data format, zero suppressed data is recorded for each pad over all the time bins.
This means, if we call $bunch$ a group of adjacent over threshold samples coming from one pad, the
signal can be represented bunch by bunch. Figure 1 shows a typical plot of time bin versus amplitude
for a given pad. Since the data is zero suppressed, it is sufficient to record three types of data,
the sample amplitude in a given bunch, the bunch length and the time gap between two consecutive bunches.
The amplitude distribution gives the energy loss spectrum with a long Landau tail and is not
illuminating for the present purpose. Therefore, in the following, we will consider only the time gap
and bunch length distributions built over all the pads.

The time gap distribution corresponds to the distance between two trajectories along the drift
direction (say z-direction). Figure 2 shows the plot of time gap distribution at different
multiplicities ($M=20000$, $40000$, $60000$ and $80000$). Although, the tail of the distribution
is linear (in the semi-log scale) and depends on $M$ as expected, it deviates from the linearity at the shorter distances. The
above behavior can be described by a distribution of the type,
\equation f(z)=A z^{-\beta} e^{-\alpha z}~~~~~~for~~~z \ge 1 \endequation
The constant $A$ is fixed by the requirement of normalization which gives $A=[Li_\beta(e^{-\alpha})]^{-1}$
where $Li_n(x)$ is the $n^{th}$ polylogarithm of $x$ 
given by,
\equation Li_n(x)=\sum_{i=1}^\infty \frac{x^i}{i^n} \endequation
Defining $\bar z = \sum_{z=1}^\infty zf(z)/\sum_{z=1}^\infty f(z)$, the average of the above distribution is
\equation \bar z = \frac{Li_{\beta-1}(e^{-\alpha})}{Li_{\beta}(e^{-\alpha})} \endequation
Note that for $\beta \rightarrow 0$, the above distribution becomes a pure exponential with 
$\bar z=(1-e^{-\alpha})^{-1}$.

\begin{figure}
\centerline{\hbox{
\psfig{figure=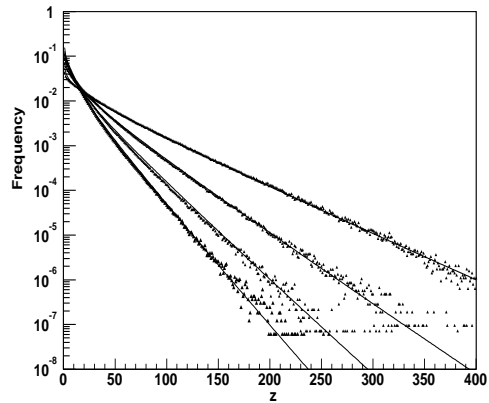,width=3.0in,height=2.6in}}}
\caption{
The frequency versus time gap for $M=20000$, $40000$, $60000$ and
$80000$ (top to bottom). The solid curves are the best fit 
obtained using Eq.(1) with different $\alpha$ and $\beta$
values as shown
in figure 3.}
\end{figure}

Figure 3 shows the parameters $\beta$ (filled circles in the left) and $\alpha$ (filled circles on the right)
extracted from fitting the simulated data points. Note that $\alpha$ is a statistical parameter which
has a near linear dependence on the multiplicty $M$. However, the index $\beta$ increases with $M$ and becomes  
a (nearly) constant at higer $M$ values. We will show below that this index is responsible for the dynamical
scaling which does not depend on external parameters like multiplicity $M$ or magnetic field
 $B$ above a critical value but depends on 
certain intrinsic TPC parameters. Another interesting observation is that the product $\omega=\bar z \alpha$ also
changes very slowly with $M$ above a critical value of $M \sim 30000$. In order to appreciate the effect of 
dynamical scaling, we can plot figure 2 in a reduced scale by dividing $z$ by $\bar z$ and multiplying $f$ by
$\bar z/A$ in Eq.(1). By this rescaling, the index $\beta$ will not change, but $\alpha$ will be rescaled to
say $\omega=\alpha \bar z$. Since $\beta$ and $\omega$ are (approximately) independent of $M$ above a certain value,
the rescaled distributions will merge with each other as shown in figure 4. The solid curve is the fit of the type
$f(z) \sim z^{-\beta} e^{-\omega z}$ with $\beta \sim 0.45$ and $\omega \sim 0.67$.  Recall here that this scaling is
quite similar to the famous KNO scaling \cite{kno} which says that at high energies $s$, the probability distributions
$P_n(s)$ of producing $n$ particles in a certain collision process should exhibit the scaling relation
$\bar n P_n(s)=f(n/\bar n) $. This scaling hypothesis asserts that if we rescale $P_n(s)$ measured at different
energies via streching (shrinking) the vertical (horizontal) axes by $\bar n$, the rescaled curves will coincide
with each other. In the present case, the observed scaling is identical to KNO scaling if we replace energy $s$
by the multiplicity $M$ and $ \bar n$ by average distance $\bar z$ although both corresponds to two different
physical situations.

\begin{figure}
\centerline{\hbox{
\psfig{figure=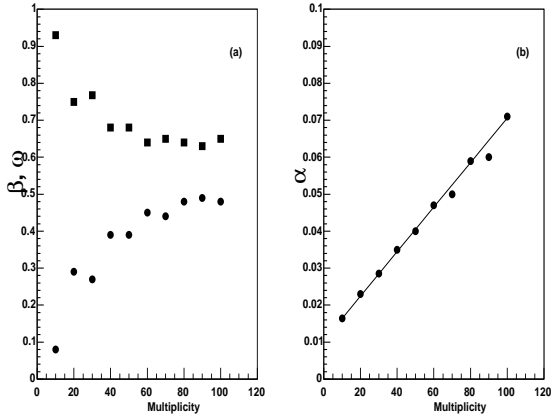,width=3.0in,height=2.6in}}}
\caption{
The dynamical index $\beta$ and $\omega={\bar z}\alpha$ (a) and $\alpha$ (b)
as a function of
Multiplicities in units of 1000. The solid line is the linaer 
fit $\alpha=0.000000602 M+ 0.0104$.}
\end{figure}

Apart from this striking similarity with KNO, 
this rescaling is an elegant way to remove the statistical dependency from the dynamical behavior.
It may be mentioned here that such type of scaling is also expected in case of an exponential distribution when
the exponent $\alpha$ is small . Under such limit ($\alpha << 1$), since the average of an exponential distribution
$\bar z = (1-e^{-\alpha})^{-1} \sim \alpha^{-1}$, $\omega=\alpha \bar z \sim 1$ (The $\omega$ value may
become more than unity
if $\alpha$ is large). However, the interesting aspect of the present scaling is the deviation
from unity and also having same $\omega$ at all $M$. This means $\omega$ deviates from unity due to presence of the dynamical exponent $\beta$. In absence of
$\beta$, the scaling would have followed the linear behaviour in the semi-log scale 
as shown by the dashed curve in figure 4. Therefore,
both $\beta$ and $\eta=1-\omega$ are dynamical exponents whose values do not depend on $M$.

In the following, we investigate the origin of this dynamical behavior and also the parameters that affect its
values. The origin of this dynamical phenomena can be associated with the set of measurements which are strongly correlated. Although,
at a given interaction point, the creation of primary electrons due to ionization is a random statistical
process, the set of interaction points created by a single charged particles passing through the TPC
are well correlated. Obviously,
this correlation will depend on the multiplicty $M$ and also on the number of measurements (number of pads, size etc).
Depending on these parameters, a critical value is reached beyond which the correlation becomes independent of $M$.

\begin{figure}
\centerline{\hbox{
\psfig{figure=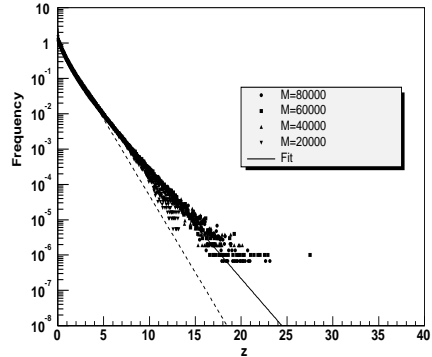,width=3.0in,height=2.6in}}}
\caption{
The frequency distribution in the reduced scale i.e. $\omega f /A$ versus $z/{\bar z}$
for different multiplicities. 
The solid line is a common fit with $\beta=0.45$ and $\omega=0.67$. The dashed line represents the 
exponential $exp(-\omega z)$ with $\omega=0.67$.}
\end{figure}

Although, the dynamical behavior may depend on other geometrical TPC parameters, in this study,
we do not intend to change any such parameters as they have been optimized based on certain
physics criteria. However, we can see the effect of other relatively softer parameters like diffusion constants, drift velocity,
noise etc which are likely to change during operation. It is noticed that out of all dynamical
parameters,  $\eta$ or $(1-\omega)$ has some dependence on $D_T$ and strong 
dependence on electronic noise. Since the produced electron clouds are broadened in transverse direction due to increased 
$D_T$, the correlation effect is also enhanced. It is found that increasing $D_T$ by a factor of two, $\omega$ reduces only from
$0.67$ to $0.62$. Further increase in $D_T$ has very little effect on $\omega$.

On the other hand,    
noise above the threshold reduces the correlation effect as it acts as an additional
random source of electrons. As shown in figure 5, the middle curve is obtained with the parameters as given 
in the tablei (reffered as default parameters), while
the upper curve corresponds to $D_T$ twice the default value and the lower curve is due to an increased noise level from $1000~ e$
to $2000~e$. The correlation effect is lost due to noisy signal for which $\omega \rightarrow 1$. With increased noise level
$\omega$ rises above unity (slowly) which is typical of an exponential behavior. Note that the dynamical parameters
like $D_T$, $D_L$ and drift velocity etc are expected to vary within $< 4\%$ during operation. This small fluctuation
does not effect the $\omega$ value. On the otherhand, noise is quite unpredictable and also may go up due to the exposure 
of the electronics to
the radiation for an extended period. Therefore, the parameter $\omega$ can serve as an excellent online tool to monitor the
noise level.

\begin{figure}
\centerline{\hbox{
\psfig{figure=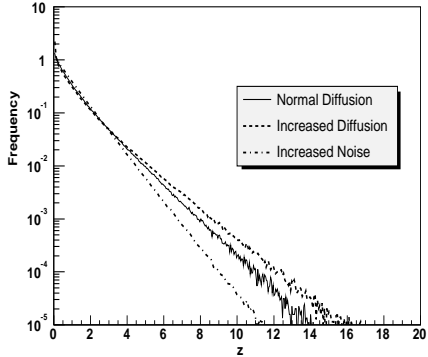,width=3.0in,height=2.6in}}}
\caption{
(a) Same as figure 4. The middle curve is with default parameters as given
in the table. The upper curve is with $D_T$ value increased by twice the default value 
where as the lower curve corresponds to
increased noise level from $1000e$ to $2000e$ for which $\omega \rightarrow 1$.}
\end{figure}

So far, we have discussed only about the time gap distribution.  The tail of the bunch length distribution
also shows a powerlaw behaviour $ \sim A~n^{-\gamma}$ as shown in figure 6(a). 
 Since the tail of the bunch length distribution
corresponds to low energy electrons and delta rays, the exponent $\gamma$ is quite sensitive to the applied
magnetic field (up to some limit).  Figure 6(b) shows the bunch length distributions at different $B$ values.
May be this exponent can be used to monitor the magnetic field setting
during the operation. Apart from this, we have not found the dependency of $\gamma$ on any other
dynamical TPC parameters which could have been utilized for online monitoring.

We would like to add here that the powerlaw exponent that affect the 
time gap distribution corresponds to a region of short distances. Therefore, it is very
difficult to extract the $\beta$ parameters acurately and also fitting the time gap distribution
with other functional forms like two exponentials
can not be ruled out. However, we have seen that the quality of fit is much better with powerlaw exponent.
Further,  as we have argued before, dynamical phenomena is expected in side a TPC 
due to many correlated interaction points.
The tail of the bunch length distribution having a perfect powerlaw behavior reflects this aspect rather unambiguously.
Due to the same dynamical origin, it is reasonable to assume a powerlaw with exponential cutoff for time gap
distribution as well.
We would also like to add
here that the KNO type of scaling found in case of time gap distribution 
is independent of any fitting procedure and the deviation of
the slope from unity is an indication that the noise level has remained below the
threshold. This we consider as an important observation of this analyses irrespective of the fitting procedures.
  
\begin{figure}
\centerline{\hbox{
\psfig{figure=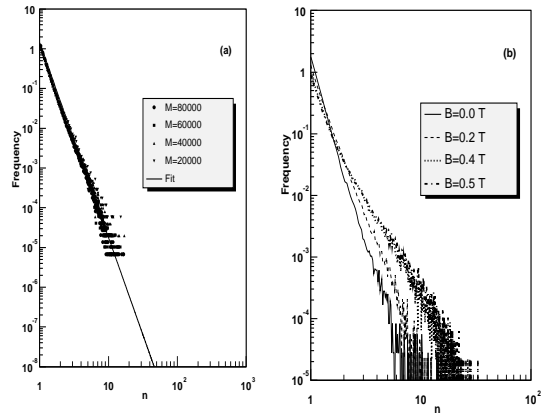,width=3.0in,height=2.6in}}}
\caption{
(a)
The bunch length distribution in the 
reduced scale for different $M$ values. The 
solid curve is a powerlaw fit with 
$\gamma \sim 4.8$. (b) Same as (a), but for different $B$ values.} 
\end{figure}

In conclusion, the time gap distribution shows an universal scaling behavior (KNO type) (at sufficiently large
multiplicity) with an exponent deviating
from unity. This is an interesting aspect to study the correlation phenomena 
in gas detector and also to understand physics (more explicitly, the physics that has gone into simulation)
at the device level. An important pratical utility of this phenomena is the utilization of the above scaling exponent to monitor
the noise level above a given threshold without any computational complexity and also without building any rigorous models.
In that sense, this analyses provide a  model independent way to monitor the quality of the data what is being recorded.

%\acknowledgements
%We would like to thank J. Belikov, M. Ivanov,
%J. Revol, J. Schukraft, and G. Valenti for useful comments and suggestions.

\begin{table}
\squeezetable
\caption{A few typical parameters used in the simulation. More details are
given in ref [5].}
\begin{tabular}{|c|c|}
Parametrs &Value\\
\hline
Diffusion Constants $(D_L=D_T)$ & 220 $\mu m /\sqrt{cm}$ \\
\hline
Drift Velocity at 400 V/cm & 2.83 $cm/\mu s$ \\
\hline
Shaping Time & 190 ns FWHM \\
\hline
Sampling Time & 200 ns \\
\hline
Noise & 1000 e \\
\hline
Magnetic field & 0.2 T \\
\hline
Oxygen Content & 5 ppm\\
\end{tabular}
\end{table}

\end{document}